\documentclass{PoS}

\newcommand{\met}       {\mbox{$\not\!\!E_T$}}

\newcommand{\ttbar}     {\mbox{$t\bar{t}$}}
\newcommand{\ppbar}     {\mbox{$p\bar{p}$}}
\newcommand{\invfb}     {\mbox{fb$^{-1}$}}

\title{Single top quark production cross section at the Tevatron}

\ShortTitle{Single top quark production cross section at the Tevatron}

\author{\speaker{Ar\'an GARC\'IA-BELLIDO}%
        University of Rochester\\
        E-mail: \email{aran.garcia-bellido@cern.ch}}

\abstract{The latest results on the measurements of electroweak top-quark production at the Tevatron are presented with the full RunII dataset. The CDF and D0 collaboration have performed measurements of the $s$, $t$, and $s+t$ channels in both $\ell$+jets and $\met$+jets final states. Evidence for the production of the $s$ channel is presented for the first time.}

\FullConference{The European Physical Society Conference on High Energy Physics \\
		 18-24 July, 2013\\
		 Stockholm, Sweden}

\begin{document}

\section{Introduction}
Top quarks are produced at the Tevatron ($\sqrt{s}=1.96$~TeV) predominantly in pairs via the strong interaction with a cross section of $\sigma(\ppbar \to \ttbar) = 7.60 \pm 0.41$~pb~\cite{tevttxs} for a top quark mass of $172.5$~GeV. In addition, top quarks are produced via the electroweak interaction in three distinct modes~\cite{Willenbrock:1986cr,Cortese:1991fw,singletop-xsec-kidonakis}: i) the $s$-channel with $\sigma(\ppbar \to tb+X) = 1.04 \pm 0.04$~pb; ii) the $t$-channel with $\sigma(\ppbar \to tqb+X) = 2.26 \pm 0.12$~pb; and iii) the associated $tW$ mode with $\sigma(\ppbar \to tW+X) = 0.28 \pm 0.06$~pb. The combined $s+t$ mode was observed for the first time in 2009 independently by CDF and D0~\cite{stop-obs-2009-cdf,stop-obs-2009-d0}, and in 2011 the D0 Collaboration measured the $t$-channel production alone with more than 5 standard deviations (SD)~\cite{Abazov:2011rz}. The $s$-channel measurement is more difficult due to its small cross section, the large irreducible background from $Wb\bar{b}$, and the lack of kinematic signatures like the forward emission of the light jet in the $t$-channel. At the LHC, the $t$-channel was observed rather quickly in 2011 by ATLAS and CMS~\cite{atlas-t,cms-t}, and the $tW$ channel in 2013 by CMS (announced in this conference). However, the $s$-channel remains a very challenging measurement at the LHC given that it requires an anti-quark in the initial state to fuse with another quark into a $W$~boson. The $t$-channel cross section is $41$ times bigger than at the Tevatron, and for $\ttbar$ it is $31$ times bigger, but the $s$-channel cross section is only $5$ times bigger than at the Tevatron. This makes the $s$-channel measurement at the Tevatron one of its legacy measurements, and one that has a better signal to background ratio at the Tevatron than at the LHC. 

The measurement of the single top quark cross section offers the possibility to probe the Standard Model (SM) structure by testing whether the $Wtb$ coupling, $|V_{tb}|$ in the CKM quark-mixing matrix, is smaller than 1, without having to assume the unitarity or three generations of the mixing matrix. The single top cross section is directly proportional to $|V_{tb}|^2$ so we can access $V_{tb}$ at production and not assume much about the CKM matrix, as is necessary in the measurement of $V_{bt}$ from the decays in $\ttbar$ production~\cite{Abazov:2011zk}. 

Finally, it is important to measure the $s$-channel and the $t$-channel independently, as new physics signatures affect their rates differently~\cite{Tait:2000sh}. This note describes the latest measurements from CDF and D0 on the measurements of the $s$, $t$, and $s+t$ productions modes using the full RunII data from the Tevatron. 

\section{D0 measurement with 9.7~\invfb~in $\ell$+jets}
D0 has optimized the single top analysis from Ref.~\cite{Abazov:2011rz} to enhance the sensitivity to the $s$-channel in the $\ell\nu bb$ final state, and has improved the selection by employing a newer more efficient $b$-tagging algorithm. The new analysis~\cite{d0-9.7ifb} now uses an inclusive trigger, the selection requires jets only up to $|\eta|<2.5$, and the events are split into four independent channels depending on their exclusive jet multiplicity (two and three jets) and the number of $b$-tagged jets (one and two or more). The dominant $W$+jets and multijet backgrounds are normalized to data before $b$-tagging, and the fractions of $W$+heavy flavor jets ($Wb\bar{b}$, $Wc\bar{c}$, and $Wcj$) to the total $W$+jets are obtained from MCFM NLO calculations. These fractions carry the largest relative systematic uncertainty with a value of 20\%. The $W$+heavy flavor normalization is cross-checked in independent samples, by looking at zero-tagged events, and by fitting the output of the $b$-tagging identification in the different $b$-tagged regions: in all cases, good agreement is found with the data and all derived scale factors by the different methods are within their statistical uncertainties. The description of the data after $b$-tagging is also checked in background dominated regions enriched in $W$+jets and, separately, in $\ttbar$, which are the two dominant backgrounds. 

Three multivariate methods are applied over the selected data: Boosted Decision Trees (BDT), Bayesian Neural Networks (BNN), and Matrix Element probability calculations (ME). They are optimized to measure the $s$ and the $t$ channels separately in all four analysis channels. The $s$-channel is the largest signal in the 2-jet, 2-$b$-tag channel; the $t$-channel is dominant in the 2 and 3-jet, 1-$b$-tag channels, and the 3-jet, 2-$b$-tag channel is mostly used to constrain the backgrounds. The BDT method uses up to 30 kinematic variables, the BNN uses the objects' four momenta with a few other variables such as the $b$-tag output of jets, $q_{\ell} \times \eta$(light jet), and the invariant $W$~boson mass, and the ME uses the objects' four vectors. The three methods are not fully correlated ($\sim 75\%$) and are their discriminant outputs are therefore combined by a second BNN which gives a better expected sensitivity than any individual method. 

\begin{figure}[!htp]
\begin{center}
\includegraphics[width=0.32\textwidth]{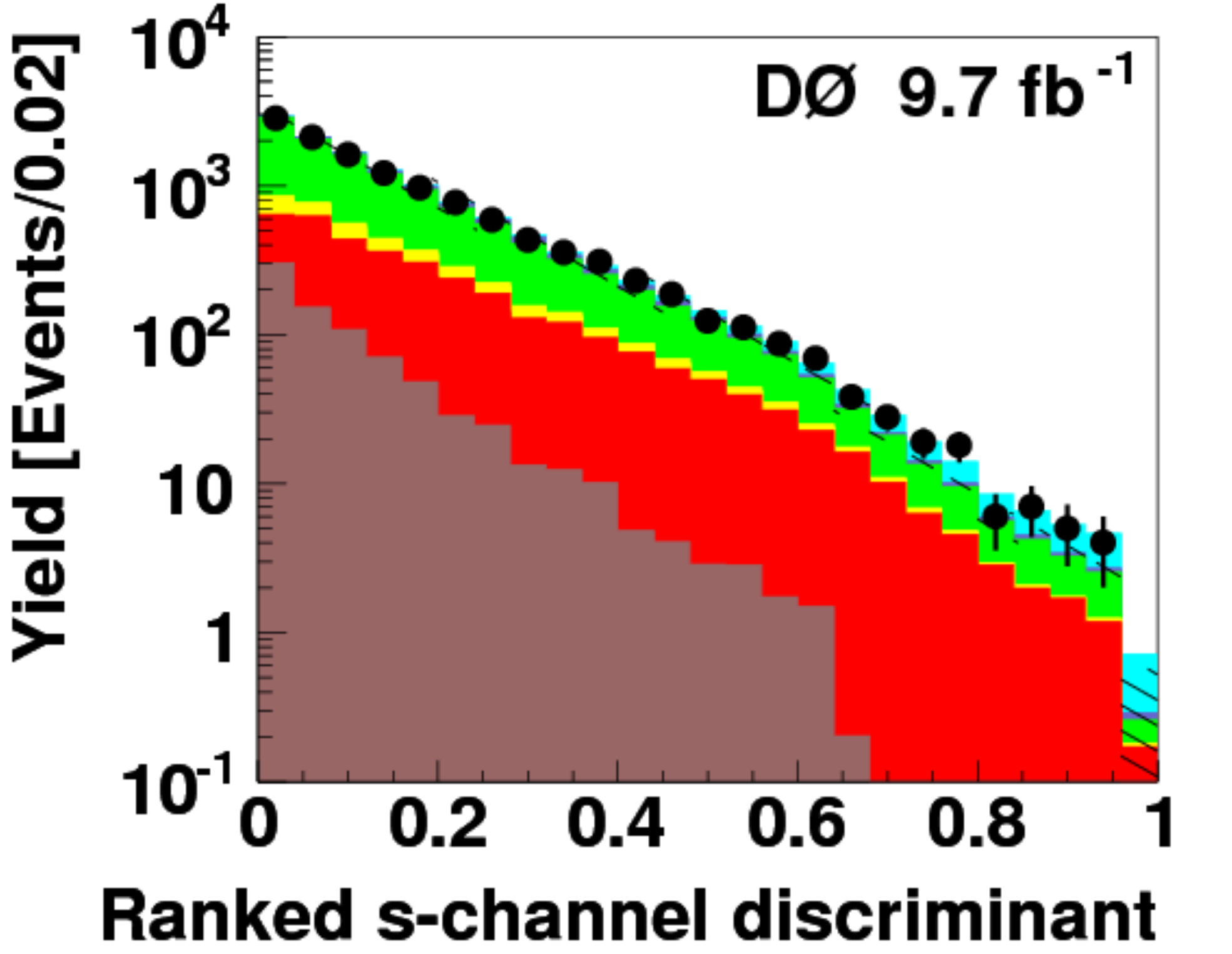}
\includegraphics[width=0.32\textwidth]{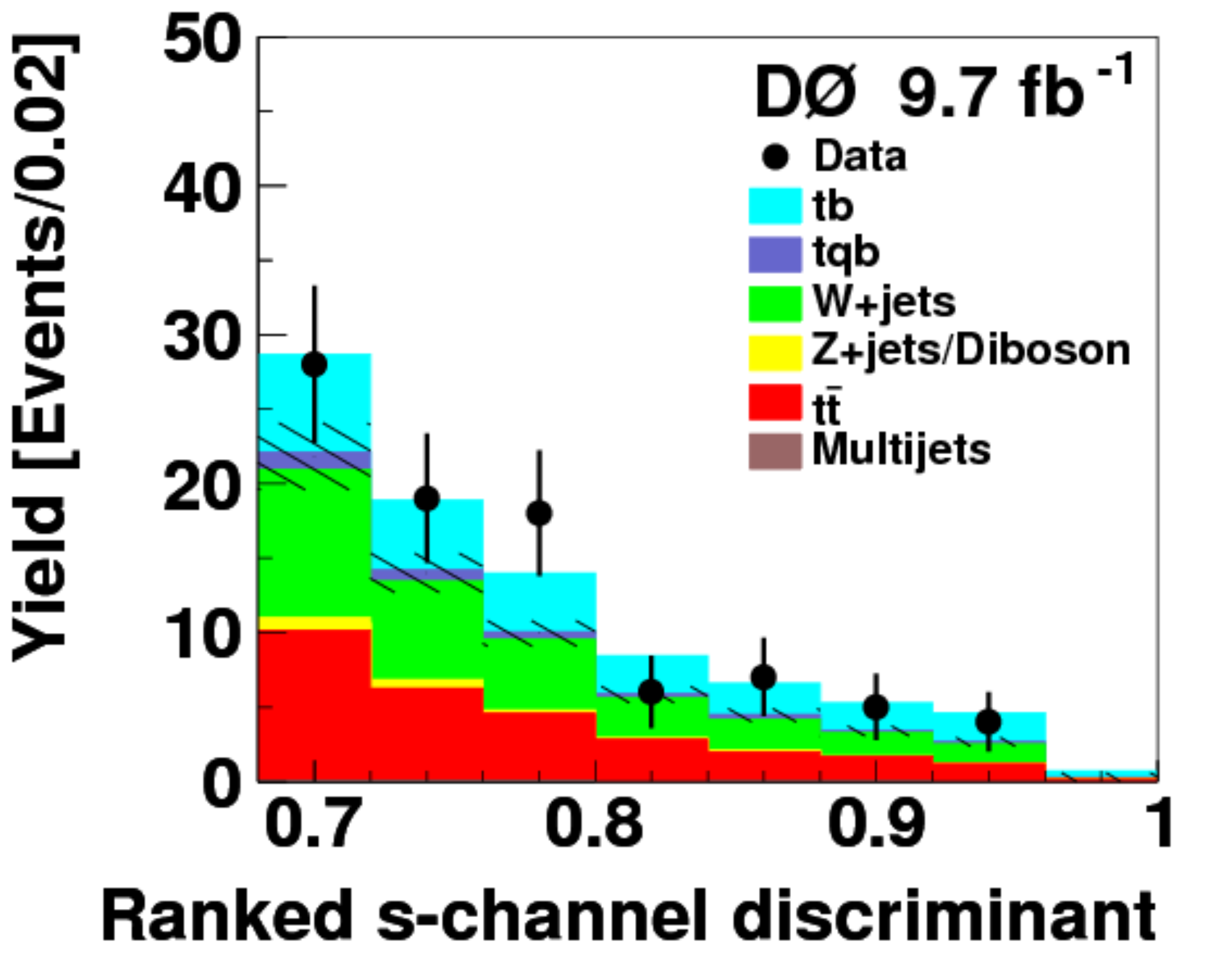}
\includegraphics[width=0.32\textwidth]{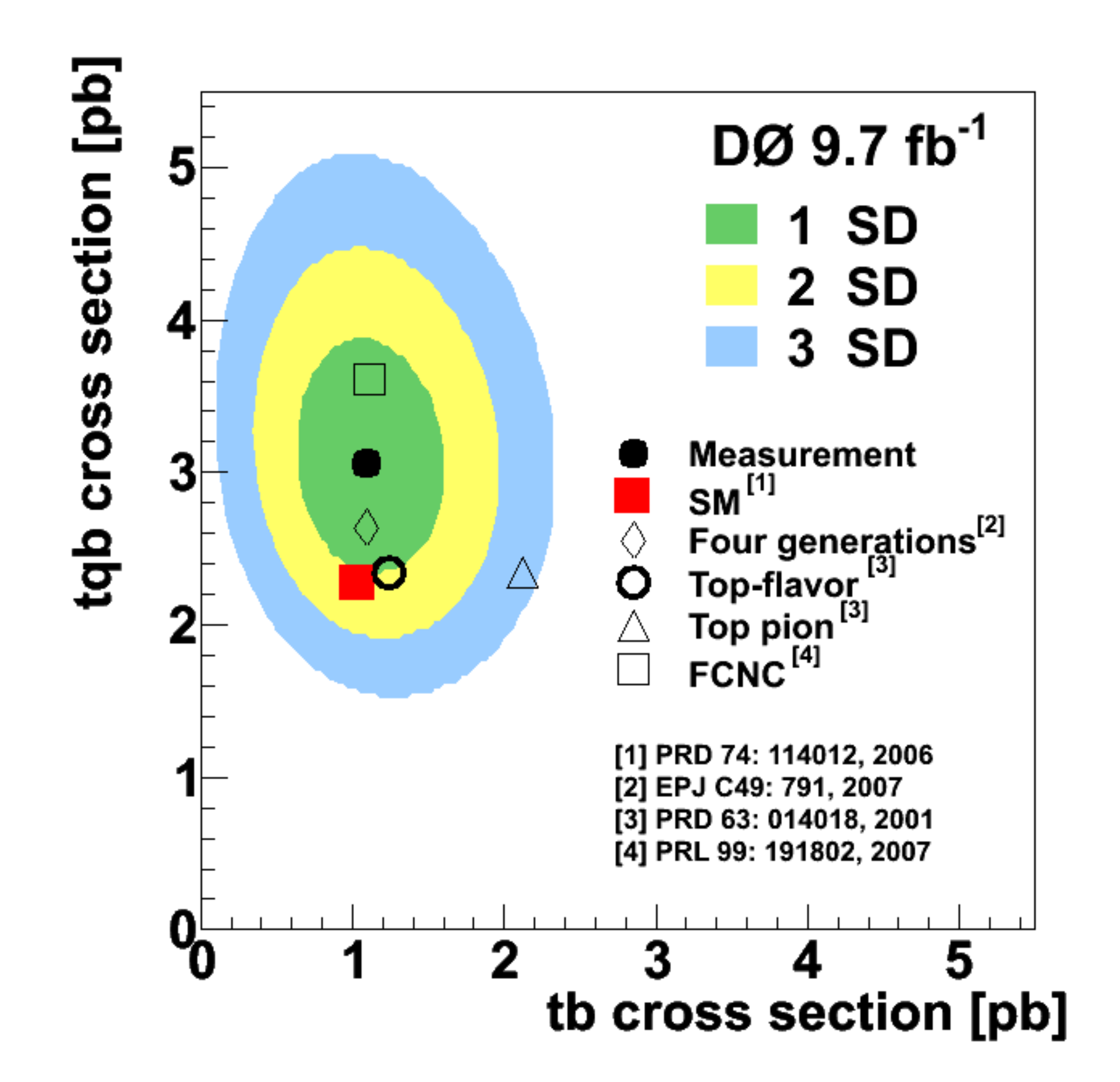}
\end{center}
\vspace{-0.5cm}
\caption[]{(left) The $s$-channel combined BNN output for all four channels in the D0 measurement. (center) Zoom of the high output discriminant, showing the measured contribution for $s$-channel (tb) events above the uncertainty on the background yield (hashed band). (right) The final two-dimensional posterior probability density as a function of the $s$-channel and the $t$-channel cross sections. Several models beyond the SM are shown for reference.\label{fig_d0}}
\end{figure}

One of the most important improvements in this analysis is the development of a new method to measure the $s$- and the $t$-channels independently, without assuming the SM value of the cross section for the other or the SM $s/t$ ratio when calculating the $s+t$ cross section. The results from the combined BNN are: $\sigma({\ppbar}{\to}tb+X) = 1.10^{+0.33}_{-0.31}$~pb,
$\sigma({\ppbar}{\to}tqb+X) = 3.07^{+0.54}_{-0.49}$~pb, and $\sigma({\ppbar}{\to}tb+tqb+X) = 4.11^{+0.60}_{-0.55}$~pb. The three different methods yield results that are in agreement with each other and also with the SM prediction. The $s$ channel is measured for the first time with a significance greater than 3~SD: the expected and observed significances are 3.7~SD. From the combined $s+t$ cross section, D0 calculates the value of $|V_{tb}|$ without any assumption on the number of generations or the unitarity of the CKM matrix, and the result is: $|V_{tb}f^L_1| = 1.12^{+0.09}_{-0.08}$, or $|V_{tb}| > 0.92$ at $95\%$~C.L., assuming a flat prior within $0 \leq |V_{tb}|^2 \leq 1$. This is currently the most stringent lower limit on $|V_{tb}|$ from the single top measurements at the Tevatron or the LHC. Figure~\ref{fig_d0} shows the D0 results.

\section{CDF measurement of the s-channel with 9.4~\invfb~in $\ell$+jets}
The CDF collaboration has presented preliminary results on the measurement of the $s$-channel cross section with the full RunII dataset in the $\ell+jets$ final state~\cite{cdf-9.4ifb-ljets}. The analysis is based on the SM Higgs boson search with 9.4~\invfb~\cite{cdf_WH}, which shares the same final state: $\ppbar \to WH \to \ell\nu b\bar{b}$. There are three lepton categories: central tight electrons, central tight muons, and extended muons (loose muons and isolated track lepton-candidates). The identification of $b$ jets has also improved with respect to previous searches and now uses the Higgs-optimized HOBIT algorithm~\cite{hobit}. The selection requires exactly two jets, at least one of which is required to be $b$-tagged. The analysis is then split in four channels according to the different $b$-tagged properties of the jets. One operational point (denoted T for tight) has 42\% $b$-jet efficiency and 0.9\% mistag rate,
and a second operational point (denoted L for loose) has 70\% $b$-jet efficiency and 9\% misidentification rate per jet. Events are therefore split according to the four orthogonal categories: TT, TL, T, and LL. This improves the sensitivity of the analysis by separating regions of the phase space with different signal-to-background ratios. 

The $W$+jets and multijet normalizations are obtained by fitting templates in the $\met$ distribution before $b$-tagging. The $W$+heavy flavor sample is derived by applying the $b$-tagging efficiency to the total pretag $W$+jets sample. The $W$+light jets (or mistags) sample is derived by applying the mistag rate parametrization to the same $W$+jets pretag sample. Multijet backgrounds are further suppressed by applying a dedicated multivariate technique. CDF includes the expected yield from the Higgs boson production (around 6 events in the TT category, for a total of 466 data events), and also includes the associated $tW$ production together with the $t$-channel as background, which D0 assumes negligible.  

The final discrimination to extract the amount of $s$-channel signal in the data is performed by training a Neural Network with the same 7 variables in each tagging category. The variables used are $M_{\ell\nu b}$, $M_{\ell\nu bb}$, $p_T(\ell)$, $M_{jj}$, $\cos\theta_{\ell j}$, $H_T$, $M^T_{\ell\nu b}$ (only for extended muons), and the $b$-jet selector output (only in the tight lepton channels).

The final result yields a value of $\sigma({\ppbar}{\to}tb+X)=1.41^{+0.44}_{-0.42}$~pb for a top quark mass of $172.5$~GeV, which corresponds to an expected sensitivity of 2.9~SD and an observed sensitivity of 3.8~SD. This result is in agreement with the SM value. CDF assumes the SM $t$-channel cross section for this measurement and does not provide a value for $|V_{tb}|$. Figure~\ref{fig_cdf} shows the CDF results.

\section{CDF measurement of the s-channel with 9.5~\invfb~in $\met$+jets}
CDF has presented preliminary results on the measurement of the $s$-channel cross section in final states with no reconstructed lepton, large missing energy, and two or three jets, with at least one $b$-tagged jet~\cite{cdf-9.5ifb-s-metjets}. This result expands a previous analysis based on the same final state that measured the combined $s+t$ channels~\cite{cdf-9.5ifb-st-metjets}. The final state and data sample is orthogonal with the $\ell$+jets analyses described above, and it is focused on events where the lepton was misreconstructed or lost, and $W$ decays to $\tau$ where the hadronic $\tau$ is reconstructed as a jet in the calorimeter. 
Similarly to the CDF $\ell$+jets analysis, the sample is divided into three categories based on the HOBIT output of the leading two jets: TT, TL, and T. The multijet background is obtained from a control region with $\met<70$~GeV and where the $\met$ and the second jet are within 0.4 radians of each other. The amount of multijet after $b$-tagging is derived from applying scaling factors in each $b$-tag category based on probability density functions from the ratio of tagged to pretagged events in several variables. The $W$+jets background is allowed to float together with the signal in the final fit to the data. 

A series of three Neural Networks are built to separate consecutively against multijet backgrounds, $\ttbar$ backgrounds and $W$+jets backgrounds. Several control regions are checked to test the validity of the background modeling after each step, and scale factors are applied as needed for each tagging category. A final discriminant output for each event is built from the quadrature weighted sum of the outputs of the $W$+jets and $\ttbar$ Neural Networks. 

\begin{figure}[!htp]
\begin{center}
\includegraphics[width=0.32\textwidth]{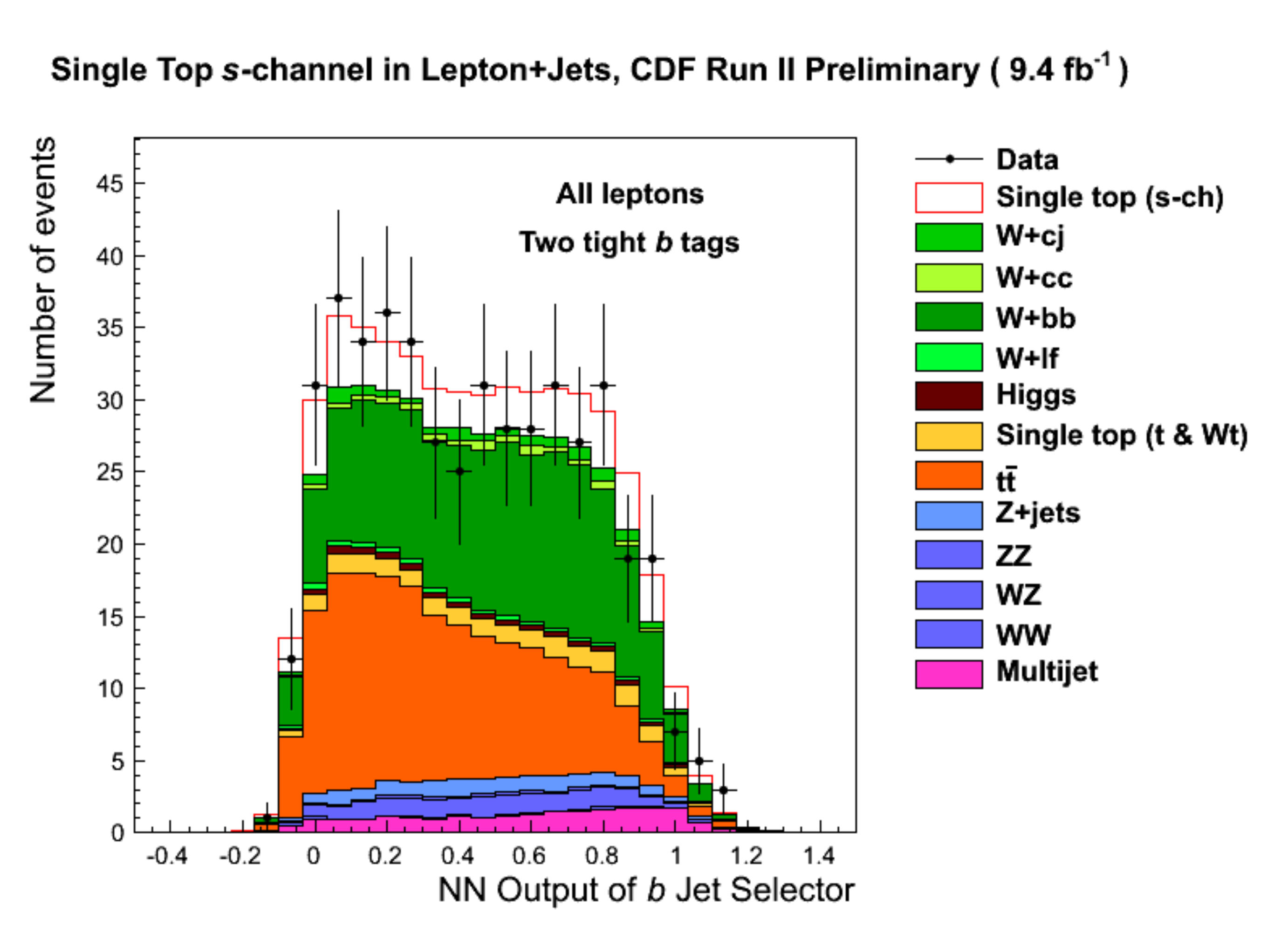}
\includegraphics[width=0.32\textwidth]{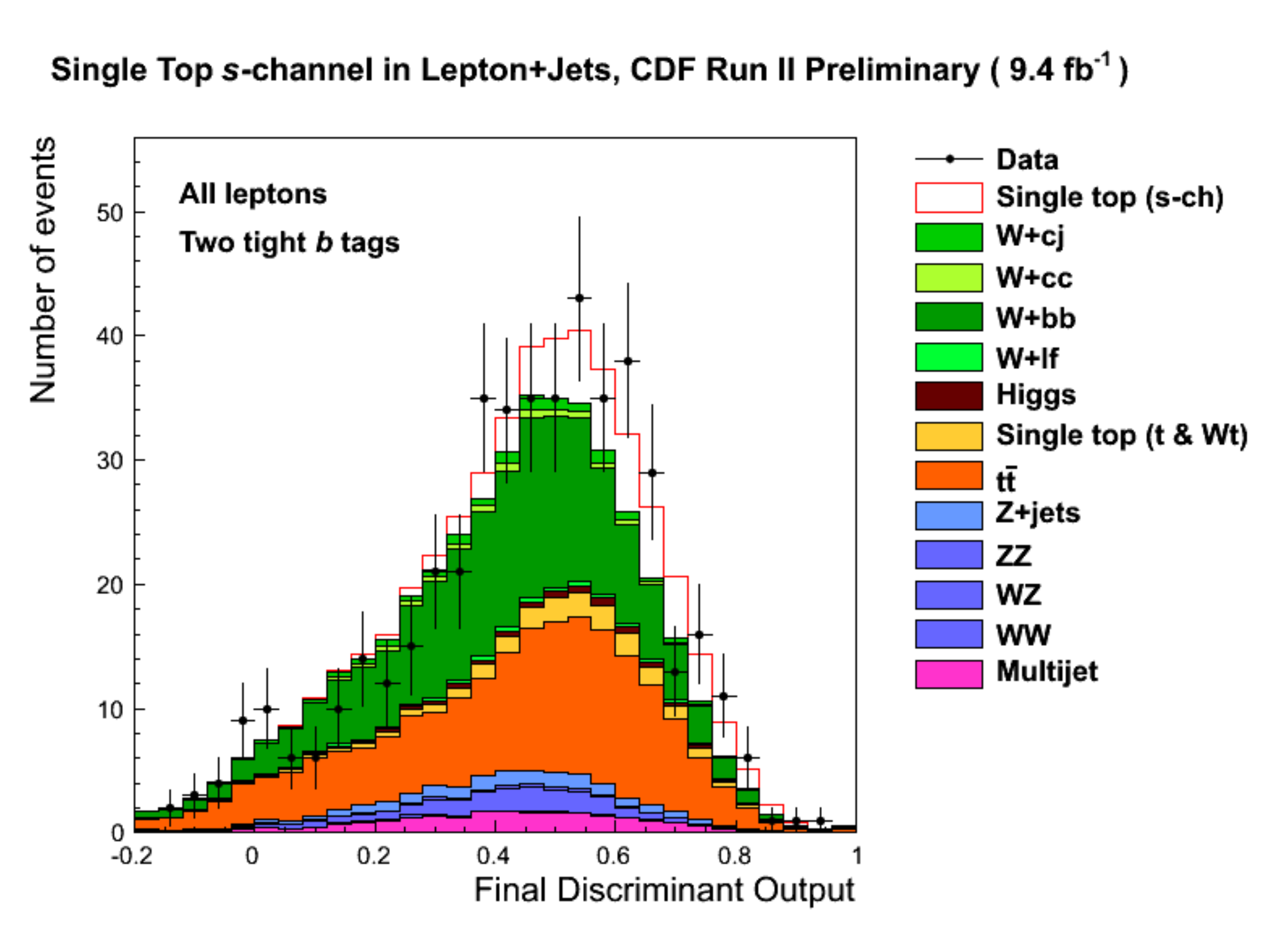}
\includegraphics[width=0.32\textwidth]{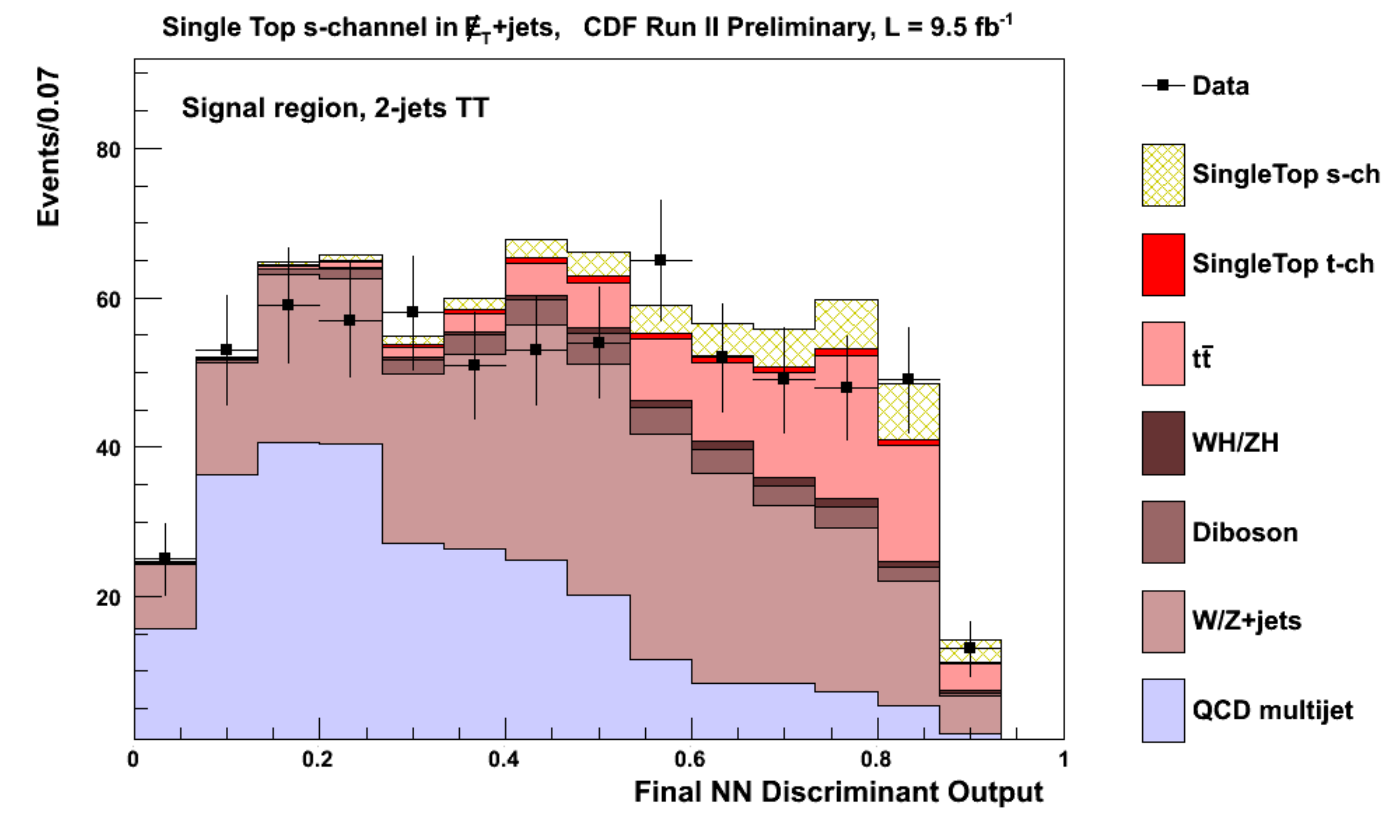}
\end{center}
\caption[]{(left) The output of the $b$-jet selector in the TT sample, which is used as input to the final NN discriminant in the $\ell$+jets analysis. (center) The output of the final NN discriminant in the $\ell$+jets TT sample. (right) The output of the final NN discriminant on the $\met$+jets TT sample.\label{fig_cdf}}
\end{figure}

The final result yields a value of $\sigma({\ppbar}{\to}tb+X)=1.10^{+0.65}_{-0.66}$~pb for a top quark mass of $172.5$~GeV. Figure~\ref{fig_cdf} shows the CDF results.

\section{Conclusions}

\begin{figure}[!h!tp]
\begin{center}
\includegraphics[width=0.45\textwidth]{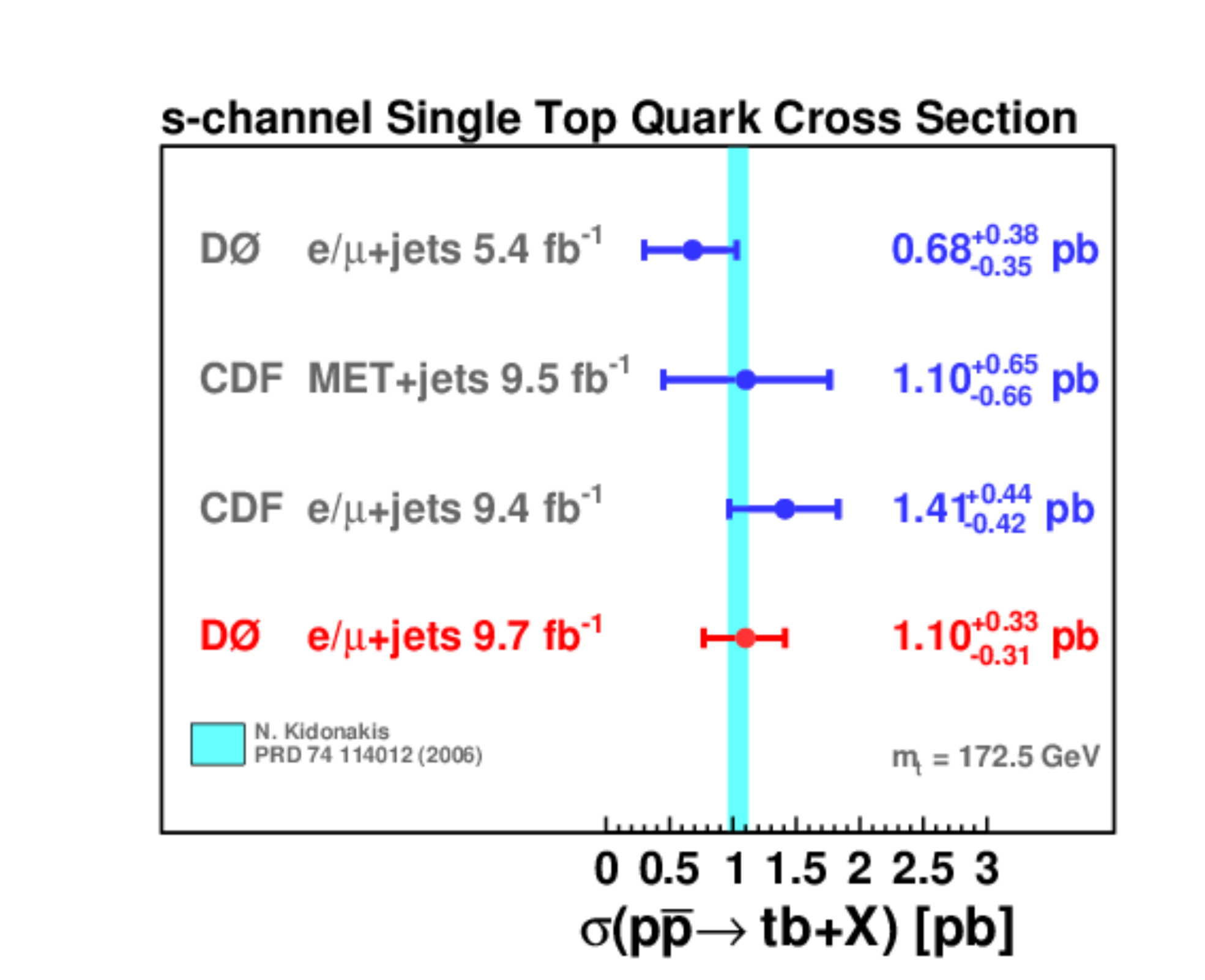}
\end{center}
\caption[]{Summary of the latest results on the $s$-channel cross section measurement at the Tevatron.\label{fig_summary}}
\end{figure}

Figure~\ref{fig_summary} shows the summary of results from all different Tevatron measurements on the $s$-channel production.  Both collaborations have measured the $s$-channel cross section with 30\% relative uncertainty and have established firm evidence for its production separately from the $t$-channel. The D0 collaboration has measured the combined $s+t$ cross section with a relative uncertainty of 15\%, and $|V_{tb}|$ with a relative uncertainty of 8\%. All results show good agreement with the SM. CDF and D0 have started the process of combining their results into a final Tevatron measurement. 

\section*{Acknowledgements}
The author wishes to thank the organizers of EPS 2013. This work was supported in part by the US Department of Energy award DE-SC0006535.


\begin{thebibliography}{99}
  \bibitem{tevttxs} 
  T.~A.~Aaltonen {\it et al.}  [CDF and  D0 Collaborations],
  arXiv:1309.7570 [hep-ex].

  \bibitem{Willenbrock:1986cr} 
  S.~S.~D.~Willenbrock and D.~A.~Dicus,
  Phys.\ Rev.\ D {\bf 34}, 155 (1986).
  \bibitem{Cortese:1991fw} 
  S.~Cortese and R.~Petronzio,
  Phys.\ Lett.\ B {\bf 253}, 494 (1991).
  
  \bibitem{singletop-xsec-kidonakis}
N.~Kidonakis,
Phys.\ Rev.\ D {\bf 74}, 114012 (2006).
\bibitem{stop-obs-2009-cdf}
T.~Aaltonen {\it et al.} [CDF Collaboration],
Phys.\ Rev.\ Lett.\ {\bf 103}, 092002 (2009).

\bibitem{stop-obs-2009-d0}
V.~M.~Abazov {\it et al.} [D0 Collaboration],
Phys.\ Rev.\ Lett.\ {\bf 103}, 092001 (2009).

\bibitem{Abazov:2011rz} 
  V.~M.~Abazov {\it et al.}  [D0 Collaboration],
  Phys.\ Lett.\ B {\bf 705}, 313 (2011)
  [arXiv:1105.2788 [hep-ex]].

    \bibitem{atlas-t}
   ATLAS Collaboration, Measurement of the t-channel single top-quark production cross section in 0.70~\invfb~pp Collisions $\sqrt{s}$ = 7 TeV with the ATLAS detector, ATLAS-CONF-2011-101 (2011).
 
  \bibitem{cms-t}
  CMS Collaboration, Measurement of the t-channel single top quark production cross sec tion in pp collisions at $\sqrt{s}$ = 7 TeV, CMS-TOP-10-008 (2011).
  
  \bibitem{Abazov:2011zk} 
  V.~M.~Abazov {\it et al.}  [D0 Collaboration],
  Phys.\ Rev.\ Lett.\  {\bf 107}, 121802 (2011)
  [arXiv:1106.5436 [hep-ex]].
  
\bibitem{Tait:2000sh}
T.~Tait and C.-P.~Yuan,
Phys. Rev. D {\bf 63}, 014018 (2001).




  
\bibitem{d0-9.7ifb}
  V.~M.~Abazov {\it et al.}  [D0 Collaboration], Accepted for publication by Phys. Lett. B.
  arXiv:1307.0731 [hep-ex].


\bibitem{cdf-9.4ifb-ljets}
CDF Conf. Note 11025, \href{http://www-cdf.fnal.gov/physics/new/top/2013/SChannelSingleTopLepJets/ST_DPF13_public.pdf}
{http://www-cdf.fnal.gov/physics/new/top/2013/SChannelSingleTopLepJets/ST\_DPF13\_public.pdf}

\bibitem{cdf_WH}
  T.~Aaltonen {\it et al.}  [CDF Collaboration],
  Phys.\ Rev.\ Lett.\  {\bf 109}, 111804 (2012)
  [arXiv:1207.1703 [hep-ex]].

\bibitem{hobit}
CDF Note 10803, 
\href{http://www-cdf.fnal.gov/physics/new/hdg/results/HOBIT_120307/Higgs_Optimized_B-jet_Identification_Tagger_public.html}{http://www-cdf.fnal.gov/physics/new/hdg/results/HOBIT\_120307/Higgs\_Optimized\_B-jet\_Identification\_Tagger\_public.html} 

\bibitem{cdf-9.5ifb-s-metjets}
CDF Conf. Note 11015,
\href{http://www-cdf.fnal.gov/physics/new/top/2013/SChannelSingleTopMETJets/cdf11015_schannelMETbbPub.pdf}{http://www-cdf.fnal.gov/physics/new/top/2013/SChannelSingleTopMETJets/cdf11015\_schannelMETbbPub.pdf}

\bibitem{cdf-9.5ifb-st-metjets}
CDF Conf. Note 10926,
\href{http://www-cdf.fnal.gov/physics/new/top/2013/SingleTop_MET_Jets/cdf10979_SingleTopMETbbPub-1.pdf}{http://www-cdf.fnal.gov/physics/new/top/2013/SingleTop\_MET\_Jets/cdf10979\_SingleTopMETbbPub-1.pdf}

  \end{thebibliography}
\end{document}